\documentclass[12pt]{article}
\usepackage{graphicx}
\usepackage{amsmath}
\usepackage{amssymb}
\usepackage{caption2}
\begin{document}
\bibliographystyle{prsty}
\begin{center}
{\large {\bf \sc{  Analysis of the $B\rightarrow a_1(1260)$
form-factors with light-cone QCD sum rules  }}} \\[2mm]
Zhi-Gang Wang \footnote{ E-mail,wangzgyiti@yahoo.com.cn.  }    \\
 Department of Physics, North China Electric Power University, Baoding 071003, P. R. China
\end{center}

\begin{abstract}
In this article, we calculate the $B\rightarrow a_1(1260)$
form-factors $V_1(q^2)$, $V_2(q^2)$, $V_3(q^2)$ and $A(q^2)$
   with the $B$-meson light-cone QCD sum rules.  Those
    form-factors are  basic parameters in studying
   the  exclusive non-leptonic two-body decays $B\to AP$ and semi-leptonic
   decays  $B\to A l \nu_l$, $B\to A \bar{l}l$.
   Our numerical results  are consistent with  the values from the (light-cone) QCD
sum rules.   The main uncertainty comes from the parameter
$\omega_0$ (or $\lambda_B$), which determines  the shapes of  the
two-particle and three-particle light-cone distribution amplitudes
of the $B$-meson, it is of great importance to refine this
parameter.
\end{abstract}


\section{Introduction}
 The weak $B\to P,V,A$ form-factors  with
 $P=\pi,K$, $V=\rho,K^*$ and $A=a_1,K_1$ final
states are basic input parameters in studying the exclusive
semi-leptonic decays  $B\to P(V,A) l \nu_l$, $B\to P(V,A) \bar{l}l$
and radiative decays $B\to V(A) \gamma $,  they also determine the
factorizable amplitudes in the non-leptonic charmless two-body
decays $B \to PP(AP,PV,VV)$. Those decays can be used  to determine
the CKM matrix elements  and to test the standard model, however, it
is a great challenge to pin down the uncertainties of the
form-factors to obtain more precise results. The exclusive
semi-leptonic decays $B\to P(V) l \nu_l$, $B\to P(V) \bar{l}l$ and
radiative decays $B\to V \gamma $ and hadronic two-body decays $B
\to PP(PV,VV)$ have been studied extensively
\cite{Form1,Form2,Form3,Form4,Form5,Form6,Form7}, while the decays
$B \to AP,VA$ have been calculated with the QCD factorization
approach \cite{Yang0705,LCSR07,Pham06}, generalized factorization
approach \cite{Wei05,Calderon07}, etc. It is more easy to deal with
the exclusive semi-leptonic precesses than the non-leptonic
precesses, and there have been many works on the relevant
form-factors $B \to \pi $, $B\to \rho $ in determining the CKM
matrix element $V_{ub}$ \cite{B-PV1,B-PV2,B-PV3,B-PV4}. The $B \to
a_1(1260) $ form-factors have been studied with the covariant
light-front approach \cite{CLF04}, ISGW2 quark model \cite{Isgur95},
quark-meson model \cite{QM99},  QCD sum rules \cite{Aliev99},
light-cone QCD sum rules \cite{LCSR07} and perturbative QCD
\cite{Lu07} . However, the values from different theoretical
approaches differ greatly from each other.

 The BaBar Collaboration and Belle Collaboration have measured the  charmless hadronic
 decays $B^0\to a_1^\pm\pi^\mp$ \cite{BaBarAp1,BelleAp1}.
Moreover, the BaBar Collaboration has  measured the time-dependent
CP asymmetries in the decays $B^0\to a_1^\pm\pi^\mp$  with
$a_1^{\mp}\to \pi^{\mp}\pi^{\pm}\pi^{\mp}$, from the measured CP
parameters, we can determine the decay rates of $a_1^+\pi^-$ and
$a_1^-\pi^+$ respectively \cite{BaBarAp1time}. Recently,  the BaBar
Collaboration has reported the observation of the decays $B^{\pm}\to
a_1^0\pi^{\pm},a_1^{\pm}\pi^0$, $ B^+\to a_1^+K^0$ and $B^0\to a_1^-
K^+$ \cite{BaBarAp107,BaBarAp1K}. So it is interesting to re-analyze
the $B \to a_1$ form-factors with the $B$-meson light-cone QCD sum
rules \cite{KhodjamirianB07}.

In Ref.\cite{KhodjamirianB07},  the authors obtain new sum rules for
the $B\to \pi,K ,  \rho,K^*$ form-factors  from the correlation
functions expanded near the light-cone in terms of the $B$-meson
distribution amplitudes, and suggest QCD sum rules motivated models
for the  three-particle $B$-meson light-cone distribution
amplitudes, which satisfy the relations given in
Ref.\cite{Qiao2001}. In Ref.\cite{Qiao2001}, the authors derive
exact relations between the two-particle and three-particle
$B$-meson light-cone distribution amplitudes from the QCD equations
of motion and heavy-quark symmetry. The two-particle $B$-meson
light-cone distribution amplitudes have been studied  with the QCD
sum rules and renormalization group equation
\cite{Grozin1997,Qiao2003,Lange2003,Braun2004,Grozin05,Khodjamirian05,Lee05}.
Although the QCD sum rules can't be used for a direct calculation of
the distribution amplitudes, it can provide constraints which have
to be implemented within the QCD motivated models  (or
parameterizations) \cite{Braun2004}.

The $B$-meson light-cone distribution amplitudes  play an important
role in the exclusive $B$-decays, the inverse moment of the
two-particle light-cone distribution amplitude $\phi_+(\omega)$
 enters many factorization formulas (for example, see Refs.\cite{Form3,Form4}).
 However, the
 light-cone distribution
 amplitudes of the $B$-meson
 are  received relatively little attention comparing with the ones of the light
 pseudoscalar mesons and vector
 mesons, our knowledge about the nonperturbative parameters which determine
  those
light-cone distribution amplitudes is limited and  an additional
application (or estimation) based on QCD is useful.

In this article, we use the $B$-meson light-cone QCD sum rules to
study the $B \to a_1$ form-factors. The  semi-leptonic decays $B\to
A l \nu_l$ can be observed at the LHCb, where the $b\bar{b}$  pairs
will be copiously produced with the cross section about $500 \,\mu
b$.

We can also study the form-factors with the light-cone QCD sum rules
using the light-cone distribution amplitudes of the axial-vector
mesons. Recently, the twist-2 and twist-3 light-cone distribution
amplitudes of the axial-vector mesons have been  calculated  with
the QCD sum rules \cite{Yang0705LC}.

The $B$-meson light-cone QCD sum rules have given reasonable values
for the $B\to \pi,K , \rho,K^*$ form-factors \cite{KhodjamirianB07},
so it is interesting to study the $B\to a_1$ form-factors and
cross-check the properties of the $B$-meson light-cone  distribution
amplitudes. Furthermore,  it is necessary
 to investigate the form-factors with different approaches and
compare the predictions of different approaches.

The article is arranged as: in Section 2, we derive the
$B\rightarrow a_1(1260)$  form-factors      with the light-cone QCD
sum rules; in Section 3, the numerical result and discussion; and
Section 4 is reserved for  conclusion.

\section{$B\rightarrow a_1(1260)$  form-factors  with light-cone QCD sum rules}

In the following, we write down the definitions for the weak
form-factors  $V_1(q^2)$, $V_2(q^2)$, $V_3(q^2)$, $V_0(q^2)$ and
$A(q^2)$ \cite{CLF04},
\begin{eqnarray}
\langle a_1(p)|J_\mu (0)|B(P)\rangle&=&i\left\{
(M_B-M_a)\epsilon_\mu^*V_1(q^2)-\frac{\epsilon^* \cdot
P}{M_B-M_a}(P+p)_\mu V_2(q^2) \right. \nonumber\\
&&\left.-2M_a\frac{\epsilon^* \cdot P}{q^2}q_\mu[V_3(q^2)-V_0(q^2)]
\right\} \, ,\\
\langle a_1(p)|J^A_\mu(0)|B(P)\rangle&=& \frac{1}{M_B-M_a}
\epsilon^{\mu\nu\alpha\beta}\epsilon_\nu^* (P+p)_\alpha q_\beta
A(q^2) \, ,
\end{eqnarray}
where
\begin{eqnarray}
V_3(q^2)&=&\frac{M_B-M_a}{2M_a}V_1(q^2)-\frac{M_B+M_a}{2M_a}V_2(q^2)
\, , \nonumber \\
J_\mu(x)&=&\bar{d}(x)\gamma_\mu b(x)\, , \nonumber \\
J^A_\mu(x)&=&\bar{d}(x)\gamma_\mu \gamma_5 b(x)\, ,
\end{eqnarray}
$V_0(0)=V_3(0)$, and the $\epsilon_\mu$ is the polarization vector
of the axial-vector meson $a_1(1260)$.
 We study the weak form-factors  $V_1(q^2)$, $V_2(q^2)$, $V_3(q^2)$,
$V_0(q^2)$ and $A(q^2)$ with the
 two-point correlation functions $\Pi^i_{\mu}(p,q)$,
\begin{eqnarray}
\Pi^i_{\mu\nu}(p,q)&=&i \int d^4x \, e^{i p \cdot x}
\langle 0 |T\left\{J^a_\mu(x) J^i_{\mu}(0)\right\}|B(P)\rangle \, ,\nonumber \\
J^a_\mu(x)&=&\bar {u}(x)\gamma_\mu \gamma_5 d(x)\, ,
\end{eqnarray}
 where  $J^i_\mu(x)=J_\mu(x)$ and $J^A_\mu(x)$ respectively, and the
 axial-vector current $J^a_\mu(x)$ interpolates
 the axial-vector meson $a_1(1260)$.
 The correlation functions
$\Pi^i_{\mu}(p,q)$ can be decomposed as
\begin{eqnarray}
\Pi^1_{\mu}(p,q)&=&\Pi_{A}g_{\mu\nu}+\Pi_{B}q_\mu p_\nu+\Pi_{C}p_\mu
q_\nu +\Pi_{D}q_\mu q_\nu+\Pi_{D}p_\mu p_\nu\,, \nonumber \\
\Pi^2_{\mu}(p,q)&=&\Pi_{2}\epsilon_{\mu\nu\alpha\beta}p^\alpha
q^\beta+\cdots
\end{eqnarray}
due to  Lorentz covariance.  In this article, we derive the sum
rules with the tensor structures $g_{\mu\nu}$, $q_\mu p_\nu$ and
$\epsilon_{\mu\nu\alpha\beta}p^\alpha q^\beta$ respectively to avoid
 contaminations from the $\pi$ meson.

According to the basic assumption of current-hadron duality in the
QCD sum rules approach \cite{SVZ79,Reinders85}, we can insert  a
complete series of intermediate states with the same quantum numbers
as the current operator  $J^a_\mu(x)$  into the correlation
functions  $\Pi^i_{\mu}(p,q) $ to obtain the hadronic
representation. After isolating the ground state  contributions from
the pole terms of the meson $a_1(1260)$,  the correlation functions
 $\Pi^i_{\mu\nu}(p,q)$  can be expressed in the following form,
\begin{eqnarray}
\Pi^1_{\mu\nu}(p,q) &=&-\frac{if_{a}M_a(M_B-M_a)V_1(q^2)}
  {M_{a}^2-p^2}g_{\mu\nu}  +\nonumber \\
  &&\frac{2 if_{a}M_aV_2(q^2)}
  {(M_B-M_a)(M_{a}^2-p^2)}q_\mu p_\nu  + \cdots \, , \\
     \Pi^2_{\mu\nu}(p,q) &=&\frac{2 f_{a}M_a A(q^2)}
  {(M_B-M_a)(M_{a}^2-p^2)} \epsilon_{\mu\nu\alpha\beta}p^\alpha q^\beta + \cdots\, ,
 \end{eqnarray}
 where we have used the standard definition  for the  decay
 constant $f_a$, $\langle0|J^a_{\mu}(0)|a_1(p)\rangle= f_{a}M_a
 \epsilon_\mu$.

 In the following, we briefly outline
the operator product expansion for the correlation functions
$\Pi^i_\mu(p,q)$ in perturbative QCD theory. The calculations are
performed at the large space-like momentum region $p^2\ll 0$ and $0
\leq q^2< m_b^2+m_b p^2/\bar{\Lambda}$, where
$M_B=m_b+\bar{\Lambda}$ in the heavy quark limit. We write down the
propagator of a massless  quark in the external gluon field in the
Fock-Schwinger gauge and the light-cone distribution amplitudes of
the $B$ meson firstly \cite{Belyaev94},
\begin{eqnarray}
\langle 0 | T \{q_i(x_1)\, \bar{q}_j(x_2)\}| 0 \rangle &=&
 i \int\frac{d^4k}{(2\pi)^4}e^{-ik(x_1-x_2)}\nonumber\\
 &&\left\{
\frac{\not\!k }{k^2} \delta_{ij} -\int\limits_0^1 dv
 G^{ij}_{\mu\nu}(vx_1+(1-v)x_2)
 \right. \nonumber \\
&&\left. \left[ \frac12 \frac {\not\!k }{k^4}\sigma^{\mu\nu} -
\frac{1}{k^2}v(x_1-x_2)^\mu
\gamma^\nu \right]\right\}\, , \nonumber\\
 \langle 0|\bar{q}_{\alpha}(x)
h_{v\beta}(0) |B(v)\rangle &=& -\frac{if_B m_B}{4}\int\limits
_0^\infty d\omega e^{-i\omega v\cdot
x} \nonumber\\
&& \left \{(1 +\not\!v) \left [ \phi_+(\omega) -
\frac{\phi_+(\omega) -\phi_-(\omega)}{2 v\cdot x}\not\! x \right
]\gamma_5\right\}_{\beta\alpha} \, , \nonumber\\
\langle 0|\bar{q}_\alpha(x) G_{\lambda\rho}(ux)
h_{v\beta}(0)|B(v)\rangle &=& \frac{f_Bm_B}{4}\int\limits_0^\infty
d\omega \int\limits_0^\infty d\xi  e^{-i(\omega+u\xi) v\cdot x}
\nonumber \\
&& \left\{(1 + \not\!v) \left[
(v_\lambda\gamma_\rho-v_\rho\gamma_\lambda)
\Big(\Psi_A(\omega,\xi)-\Psi_V(\omega,\xi)\Big)
\right.\right.\nonumber\\
&&-i\sigma_{\lambda\rho}\Psi_V(\omega,\xi)-\frac{x_\lambda
v_\rho-x_\rho v_\lambda}{v\cdot x}X_A(\omega,\xi)\nonumber\\
&&\left.\left. +\frac{x_\lambda \gamma_\rho-x_\rho
\gamma_\lambda}{v\cdot
x}Y_A(\omega,\xi)\right]\gamma_5\right\}_{\beta\alpha}\,,
\end{eqnarray}
where
\begin{eqnarray}
\phi_+(\omega)&=&
\frac{\omega}{\omega_0^2}e^{-\frac{\omega}{\omega_0}} \, ,\,\,\,
\phi_-(\omega)= \frac{1}{\omega_0}e^{-\frac{\omega}{\omega_0}} \, ,\nonumber\\
 \Psi_A(\omega,\xi)& =& \Psi_V(\omega,\xi) = \frac{\lambda_E^2
}{6\omega_0^4}\xi^2 e^{-\frac{\omega+\xi}{\omega_0}} \, ,\nonumber\\
X_A(\omega,\xi)& = & \frac{\lambda_E^2 }{6\omega_0^4}\xi(2\omega-\xi)e^{-\frac{\omega+\xi}{\omega_0}}\,,\nonumber\\
Y_A(\omega,\xi)& =&  -\dfrac{\lambda_E^2 }{24\omega_0^4}
\xi(7\omega_0-13\omega+3\xi)e^{-\frac{\omega+\xi}{\omega_0}}\,,
\end{eqnarray}
the $\omega_0$ and $\lambda^2_E$ are some parameters of the
$B$-meson light-cone distribution amplitudes.

Substituting the $d$ quark propagator and the corresponding
$B$-meson light-cone distribution amplitudes into the correlation
functions $\Pi^i_\mu(p,q)$, and completing the integrals over the
variables $x$ and $k$, finally we obtain the representation at the
level of quark-gluon degrees of freedom.  In this article, we take
the three-particle  $B$-meson light-cone distribution amplitudes
suggested in Ref.\cite{KhodjamirianB07}, they obey the powerful
constraints derived in Ref.\cite{Qiao2001} and the relations between
the matrix elements of the local operators and the moments of the
light-cone distribution amplitudes,   if the conditions $
\omega_0=\frac{2}{3} \bar{\Lambda}$ and
$\lambda_E^2=\lambda_H^2=\frac{3}{2}\omega_0^2= \frac{2}{3}
\bar{\Lambda}^2$
 are satisfied \cite{Grozin1997}.

In the region of small $\omega$, the exponential form of
distribution amplitude $\phi_+(\omega)$ is numerically close to the
more elaborated model (or the BIK distribution amplitude (BIK DA))
suggested in Ref.\cite{Braun2004},
\begin{eqnarray}
\phi_+(\omega, \mu=1 \mbox{GeV}) =\frac{4\omega}{\pi
\lambda_B(1+\omega^2)} \left[\frac{1}{1+\omega^2}-2
\frac{\sigma_B-1}{\pi^2} \ln\omega\right]\, ,
 \end{eqnarray}
 where $\omega_0=\lambda_B$. The parameters $\lambda_B$ and
$\sigma_B$ are determined from the heavy quark effective theory QCD
sum rules including the  radiative and nonperturbative corrections.
There are other phenomenological models for the two-particle
$B$-meson light-cone distribution amplitudes, for example, the $k_T$
factorization formalism \cite{Bdis1,Bdis2}, in this article, we use
the QCD sum rules motivated models.

 After matching  with the hadronic representation below the continuum threshold $s_0$, we
obtain the following three sum rules for the weak form-factors
$V_1(q^2)$ , $V_2(q^2)$ and $A(q^2)$ respectively,
\begin{eqnarray}
V_1(q^2) &=& \frac{1}{f_a M_a(M_B-M_a)}e^{\frac{M_a^2}{M^2}}\left\{-
\frac{1}{2}f_BM_BM^2\int_{0}^{\sigma_0}d\sigma
\phi_+(\omega')\frac{d}{d\sigma}e^{-\frac{s}{M^2}} \right.\nonumber\\
&&-\frac{f_BM_B}{2}\int_{0}^{\sigma_0}d\sigma \int_0^{\sigma
M_B}d\omega \int_{\sigma M_B-\omega}^\infty
\frac{d\xi}{\xi}[\Psi_A(\omega,\xi)-\Psi_V(\omega,\xi)]\frac{d}{d\sigma}
\frac{1}{\bar{\sigma}}e^{-\frac{s}{M^2}} \nonumber \\
&&+\frac{f_BM_B^2}{M^2}\int_{0}^{\sigma_0}d\sigma \int_0^{\sigma
M_B}d\omega \int_{\sigma M_B-\omega}^\infty\frac{d\xi}{\xi}
\frac{(1-2u)[3\widetilde{X}_A(\omega,\xi)-2\widetilde{Y}_A(\omega,\xi)]}
{\bar{\sigma}^2}e^{-\frac{s}{M^2}} \nonumber \\
&&-f_B\int_{0}^{\sigma_0}d\sigma \int_0^{\sigma M_B}d\omega
\int_{\sigma M_B-\omega}^\infty\frac{d\xi}{\xi}
\frac{(1-2u)\widetilde{X}_A(\omega,\xi)}
{\bar{\sigma}^3}e^{-\frac{s}{M^2}} \nonumber \\
&&\left.\left[\frac{\widetilde{M}_B^4-4sM_B^2}{2M^4}-2\frac{\widetilde{M}_B^2-2M_B^2}{M^2}+1
\right] \right\} \, ,
\end{eqnarray}

\begin{eqnarray}
 V_2(q^2) &=& \frac{M_B-M_a}{2f_a M_a}
e^{\frac{M_a^2}{M^2}}\left\{f_BM_B\int_{0}^{\sigma_0}d\sigma \left[
\phi_+(\omega')\frac{1-2\sigma}{\bar{\sigma} } +\right. \right.\nonumber \\
&&\left.\frac{2M_B}{M^2}
[\widetilde{\phi}_+(\omega')-\widetilde{\phi}_-(\omega')]\frac{\sigma}{\bar{\sigma}}\right]e^{-\frac{s}{M^2}}\nonumber\\
&&+\frac{f_BM_B}{M^2}\int_{0}^{\sigma_0}d\sigma \int_0^{\sigma
M_B}d\omega \int_{\sigma M_B-\omega}^\infty\frac{d\xi}{\xi}
\frac{(2\sigma-3)[\Psi_A(\omega,\xi)-\Psi_V(\omega,\xi)]}
{\bar{\sigma}^2}e^{-\frac{s}{M^2}} \nonumber \\
&&+\frac{f_B}{M^2}\int_{0}^{\sigma_0}d\sigma \int_0^{\sigma
M_B}d\omega \int_{\sigma M_B-\omega}^\infty\frac{d\xi}{\xi}
(1-2u)\widetilde{X}_A(\omega,\xi)(6+\frac{d}{d\sigma})\frac{1}
{\bar{\sigma}^2}e^{-\frac{s}{M^2}} \nonumber \\
&&-\frac{4f_BM_B^2}{M^4}\int_{0}^{\sigma_0}d\sigma \int_0^{\sigma
M_B}d\omega \int_{\sigma M_B-\omega}^\infty\frac{d\xi}{\xi}
(1-2u)\widetilde{Y}_A(\omega,\xi)\frac{\sigma}
{\bar{\sigma}^2}e^{-\frac{s}{M^2}} \nonumber \\
&&-\frac{4f_B}{M^2}\int_{0}^{\sigma_0}d\sigma \int_0^{\sigma
M_B}d\omega \int_{\sigma M_B-\omega}^\infty\frac{d\xi}{\xi}
\frac{(1-2u)\widetilde{X}_A(\omega,\xi)}{\bar{\sigma}^3}\nonumber\\
&&\left.\left[2-\sigma-\frac{2s-\sigma
\widetilde{M}_B^2}{2M^2}\right]e^{-\frac{s}{M^2}} \right\}\, ,
\end{eqnarray}

\begin{eqnarray}
 A(q^2)&=&\frac{M_B-M_a}{2f_a M_a} e^{\frac{M_a^2}{M^2}}\left\{ f_B M_B
\int_{0}^{\sigma_0}d\sigma
\frac{\phi_+(\omega')}{\bar{\sigma}}e^{-\frac{s}{M^2}}  \right.\nonumber\\
&&+\frac{f_BM_B}{M^2}\int_{0}^{\sigma_0}d\sigma \int_0^{\sigma
M_B}d\omega \int_{\sigma M_B-\omega}^\infty\frac{d\xi}{\xi}
\frac{[\Psi_A(\omega,\xi)-\Psi_V(\omega,\xi)]}
{\bar{\sigma}^2}e^{-\frac{s}{M^2}} \nonumber \\
&&\left.+\frac{f_B}{M^2}\int_{0}^{\sigma_0}d\sigma \int_0^{\sigma
M_B}d\omega \int_{\sigma M_B-\omega}^\infty\frac{d\xi}{\xi}
(1-2u)\widetilde{X}_A(\omega,\xi)\frac{d}{d\sigma}\frac{1}
{\bar{\sigma}^2}e^{-\frac{s}{M^2}} \right\}\, ,
\end{eqnarray}
where
\begin{eqnarray}
s &=& M_B^2 \sigma-\frac{\sigma}{\bar{\sigma}}q^2\, , \,\,\,\omega'=\sigma M_B \,, \, \, \,\bar{\sigma}=1-\sigma \, ,\nonumber\\
\sigma_0&=&\frac{s_0+M_B^2-q^2-\sqrt{(s_0+M_B^2-q^2)^2-4s_0M_B^2}}{2M_B^2} \, ,\nonumber\\
u&=&\frac{\sigma M_B-\omega}{\xi} \, ,\,\,\,\widetilde{M}_B^2=M_B^2(1+\sigma)-\frac{1}{\bar{\sigma}}q^2\, ,\nonumber\\
\widetilde{X}_A(\omega,\xi)&=&\int_0^\omega d\lambda
X_A(\lambda,\xi) \, ,\,\,\,\widetilde{Y}_A(\omega,\xi)=\int_0^\omega
d\lambda
Y_A(\lambda,\xi) \, ,\nonumber\\
\widetilde{\phi}_\pm(\omega)&=&\int_0^\omega d\lambda
\phi_\pm(\lambda) \, .
\end{eqnarray}

In Ref.\cite{Lange2003}, Lange and  Neubert observe that the
evolution effects drive the light-cone distribution amplitude
$\phi_+(\omega)$ toward a linear growth at the origin and generate a
radiative tail that  falls off slower than $\frac{1}{\omega}$, even
if the initial function has an arbitrarily rapid falloff, which
implies the normalization integral of the $\phi_{+}(\omega)$ is
ultraviolet divergent. In this article, we derive the sum rules
without the radiative $\mathcal {O}(\alpha_s)$ corrections, the
ultraviolet behavior of the $\phi_+(\omega)$ plays no role at the
leading order ($\mathcal {O}(1)$). Furthermore, the duality
thresholds in the sum rules  are well below the region where the
effect of the tail becomes noticeable. The nontrivial
renormalization of the $B$-meson light-cone distribution amplitude
is so far known only for the $\phi_{+}(\omega)$,  we use the
light-cone distribution amplitudes of order $\mathcal {O}(1)$, which
satisfy
 all QCD constraints.

\section{Numerical result and discussion}
The input parameters  are taken as
$\omega_0=\lambda_B(\mu)=(0.46\pm0.11)\,\rm{GeV}$, $\mu=1\,\rm{GeV}$
\cite{Braun2004}, $\lambda_E^2=(0.11\pm0.06)\,\rm{GeV}^2$
\cite{Grozin1997}, $M_a=(1.23\pm0.06)\,\rm{GeV}$,
$f_a=(0.238\pm0.010)\,\rm{GeV}$, $s_0=(2.55\pm0.15)\,\rm{GeV}^2$
\cite{Yang0705LC},
 $M_B=5.279 \,\rm{GeV}$, $f_B=(0.18 \pm
0.02)\,\rm{GeV}$ \cite{LCSRreview,WangNPA}.

The Borel parameters in the three  sum rules  are taken as
$M^2=(1.1-1.5) \,\rm{GeV}^2$, in this region, the values of the weak
form-factors $V_1(q^2)$, $V_2(q^2)$ and $A(q^2)$ are  stable enough.

\begin{figure}
\centering
   \includegraphics[totalheight=6cm,width=7cm]{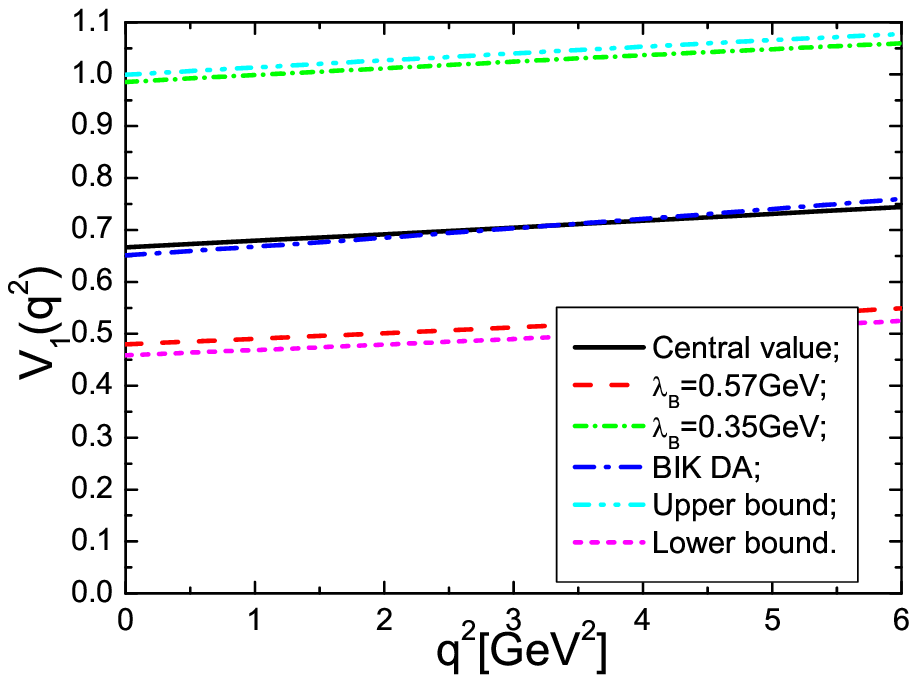}
 \includegraphics[totalheight=6cm,width=7cm]{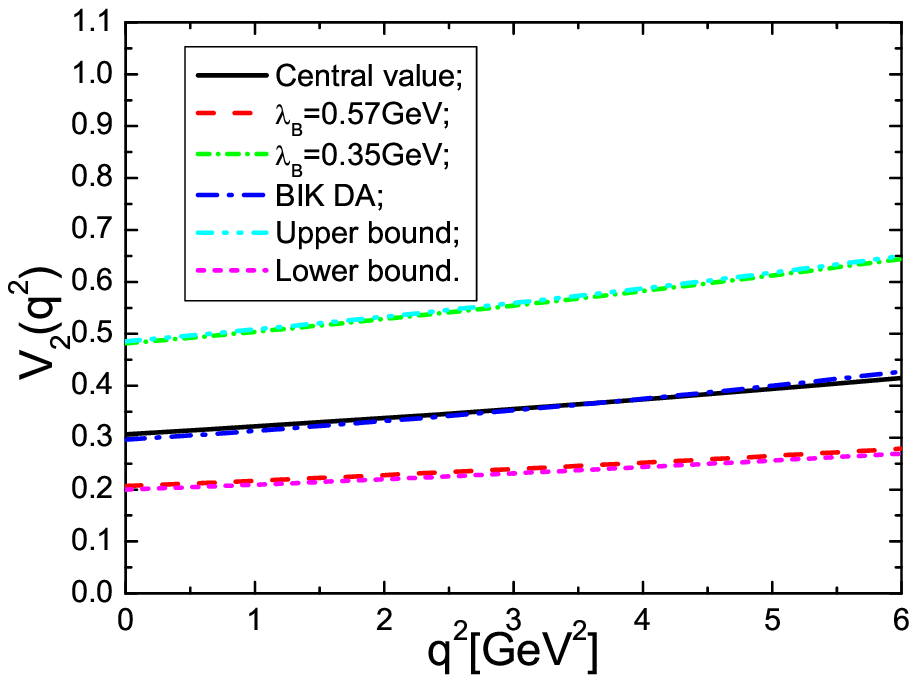}
 \includegraphics[totalheight=6cm,width=7cm]{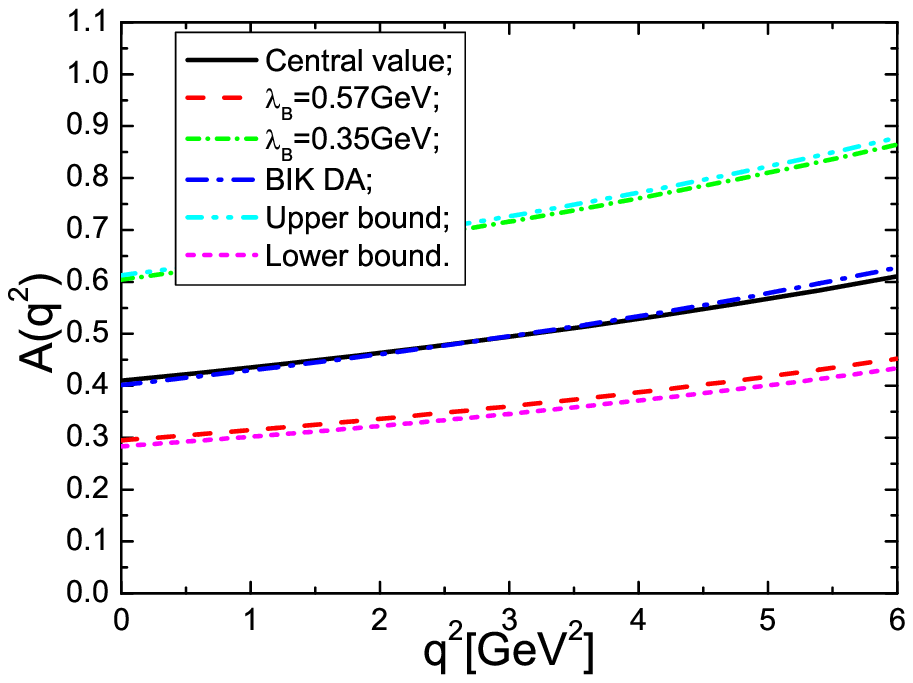}
      \caption{The form-factors $V_1(q^2)$, $V_2(q^2)$ and $A(q^2)$  with the momentum transfer $q^2$ . }
\end{figure}

\begin{table}[ht]
\centering
\begin{tabular}{|c|c|}
\hline  theoretical  approaches  & $V_0(0)$ \\
\hline
  Covariant light front approach \cite{CLF04}  & $0.13$ \\
\hline
  ISGW2 quark model \cite{Isgur95}  & $1.01$ \\
\hline
 quark-meson   model \cite{QM99}  & $1.20$ \\
\hline
QCD sum rules \cite{Aliev99} & $0.23\pm0.05$ \\
\hline
 perturbative QCD  \cite{Lu07} & $0.34^{+0.07+0.08}_{-0.06-0.08}$ \\
\hline
 light-cone sum rules \cite{LCSR07}  & $0.30\pm0.05$ \\
\hline
This work (light-cone sum rules) & $0.29^{+0.07}_{-0.06}$ \\
\hline
\end{tabular}
\caption{ The form-factor $V_0(0)$ from different theoretical
approaches. I know the updated value $0.30\pm0.05$ from private
 communication with  Prof. H.Y.Cheng, their work is still in progress.  }
\end{table}

\begin{table}[ht]
\centering
\begin{tabular}{|c|c|}
\hline  theoretical  approaches  & $A(0)$ \\
\hline
  Covariant light front approach \cite{CLF04}  & $0.25$ \\
\hline
  quark-meson   model \cite{QM99}  & $0.09$ \\
\hline
 QCD sum rules \cite{Aliev99} & $0.42\pm0.06$ \\
 \hline
  perturbative QCD  \cite{Lu07} & $0.26^{+0.06+0.03}_{-0.05-0.03}$ \\
\hline
 This work (light-cone sum rules) & $0.41^{+0.20}_{-0.13}$ \\
\hline
\end{tabular}
\caption{ The form-factor $A(0)$ from different theoretical
approaches.}
\end{table}

\begin{table}[ht]
\centering
\begin{tabular}{|c|c|c|}
\hline     & $a_F$ & $b_F$\\
\hline
  $V_1(q^2)$  & $-0.518$ & $0.159$\\
\hline
  $V_2(q^2)$  & $-1.330$ & $0.532$\\
\hline
 $A(q^2)$  & $-1.649$ & $0.561$\\
\hline
\end{tabular}
\caption{ The parameters for the fitted form-factors.}
\end{table}

Taking into account all the uncertainties, we obtain the numerical
values of the weak  form-factors $V_1(q^2)$, $V_2(q^2)$ and
$A(q^2)$, which are shown in Fig.1, at  zero momentum transfer,
\begin{eqnarray}
 V_1(0) &=&0.67^{+0.33}_{-0.21}    \, , \nonumber \\
 V_2(0) &=&0.31^{+0.18}_{-0.11}    \, , \nonumber \\
  V_3(0) &=&0.29^{+0.07}_{-0.06}    \, , \nonumber \\
 V_0(0) &=&0.29^{+0.07}_{-0.06}    \, , \nonumber \\
A(0) &=&0.41^{+0.20}_{-0.13}    \, .
 \end{eqnarray}
 The form-factors can be parameterized  in the double-pole form,
 \begin{eqnarray}
 F_i(q^2)&=&\frac{F_i(0)}{1+a_Fq^2/M_b^2+b_Fq^4/M_B^4}  \, ,
 \end{eqnarray}
where we use the notation $F_i(q^2)$ to denote the $V_1(q^2)$,
$V_2(q^2)$ and $A(q^2)$, the $a_F$ and $b_F$ are the corresponding
coefficients and their values are presented in Table 3.

In calculation, we observe  the dominating contributions in the
three sum rules come from the two-particle $B$-meson light-cone
distribution amplitudes, the contributions from the three-particle
$B$-meson light-cone distribution amplitudes are of minor
importance, about $1\%$, and can be neglected safely. It is not
un-expected that the main uncertainty comes from the parameter
$\omega_0$ (or $\lambda_B$), which determines the shapes of   the
two-particle and three-particle light-cone distribution amplitudes
of the $B$ meson. From Fig.1, we can see that the uncertainty of the
parameter $\lambda_B$ almost saturates the total uncertainties,  it
is of great importance to refine this parameter. In this article, we
take the value from the QCD sum rules in Ref.\cite{Braun2004}, where
the $B$-meson light-cone distribution amplitude $\phi_+$ is
parameterized  by the matrix element of the bilocal operator at
imaginary light-cone separation.

In the region of small $\omega$, the exponential (Gaussian) form of
distribution amplitude $\phi_+(\omega)$ is numerically close to the
BIK DA suggested in Ref.\cite{Braun2004}. In Fig.1, we also present
the numerical results with  the BIK DA  for the central values of
the input parameters $\lambda_B$ and $\sigma_B$, the Gaussian
distribution amplitude and the BIK DA lead to almost the same
values.

From  Table 1, we can see that the values of the $V_0(0)$ from
the covariant light-front approach, ISGW2 quark model and
quark-meson model differ greatly from the corresponding ones from
the (light-cone) QCD sum rules, while the values
from the (light-cone) QCD sum rules and perturbative QCD are
consistent with each other. From  Table 2, we observe that the
values of the $A(0)$ from the covariant light-front approach,
quark-meson model and perturbative QCD differ greatly from the
corresponding ones from the (light-cone) QCD sum rules, while the
values of the form-factors from the (light-cone) QCD sum rules are
consistent with each other.

\section{Conclusion}
In this article, we calculate the weak form-factors $V_1(q^2)$,
$V_2(q^2)$, $V_3(q^2)$ and $A(q^2)$
   with the $B$-meson light-cone QCD sum rules.  The form-factors are  basic parameters in studying
   the  exclusive hadronic  two-body decays  $B\to AP$  and semi-leptonic decays $B\to A l \nu_l$, $B\to A \bar{l}l$.
   Our numerical
values  are consistent with  the values from the (light-cone) QCD
sum rules.   The main uncertainty comes from the parameter
$\omega_0$ (or $\lambda_B$), which determines  the shapes of  the
two-particle and three-particle light-cone distribution amplitudes
of the $B$ meson, it is of great importance to refine this
parameter. However, it is a difficult work, as  we cannot extract
the values of the basic parameter $\lambda_B$ directly  from the
experimental data on the semi-leptonic decays $B\to A l \nu_l$.

\section*{Acknowledgments}
This  work is supported by National Natural Science Foundation,
Grant Number  10775051, and Program for New Century Excellent
Talents in University, Grant Number NCET-07-0282.

\end{document}